\documentclass[12pt]{article}
\usepackage{latexsym,graphicx}
\usepackage{subfigure}
\usepackage{amssymb}
\usepackage{amsmath}
\usepackage{amscd}
\usepackage{float}
\usepackage{amsthm}
\usepackage[left=2cm,top=2.5cm,right=2.5cm,bottom=1.5cm]{geometry}  
\usepackage{color}  
\begin{document}
\begin{center}
\large{\bf{A stable flat universe with variable Cosmological constant in $f(R,T)$ Gravity}} \\
\vspace{10mm}
\normalsize{Nasr Ahmed$^1$ and Sultan Z. Alamri$^2$}\\
\vspace{5mm}
\small{\footnotesize $^1$ Astronomy Department, National Research Institute of Astronomy and Geophysics, Helwan, Cairo, Egypt\footnote{nasr.ahmed@nriag.sci.eg}} \\
\small{\footnotesize $^2$ Mathematics Department, Faculty of Science, Taibah University, Saudi Arabia\footnote{szamri@taibahu.edu.sa}} \\
\end{center}  
\date{}
\begin{abstract}

In this paper, a general FRW cosmological model has been constructed in $f(R,T)$ gravity reconstruction with variable cosmological constant. A number of solutions to the field equations has been generated by utilizing a form for the Hubble parameter that leads to Berman's law of constant deceleration parameter $q=m-1$. The possible decelerating and accelerating solutions
have been investigated. For ($q>0$) we get a stable flat decelerating radiation-dominated universe at $q=1$. For ($q<0$) we get a stable accelerating solution describing a flat universe with positive energy density and negative cosmological constant. Nonconventional mechanisms that are expected to address the late-time acceleration with negative cosmological constant have been discussed.

\end{abstract}
\section{Introduction}

The recent observations suggest that the present universe is flat (de Bernardis et al. 2000; Bennett et al. 2003; Spergel et al. 2003) and expanding
with acceleration (Perlmutter et al. 1999; Riess et al. 1998; Percival et al. 2001; Jimenez et al. 2003). This accelerating expansion can not be explained in the framework of general relativity and the reason behind it is assumed to be an exotic form of energy with negative pressure known as ‘dark energy’ (DE). Other observations (Spergel et al. 2003; Ade et al. 2013; Eisenstein et al. 2005) confirmed that about 70 \% of the universe is made of dark energy. Several dynamical dark energy models with the rolling scalar fields have been proposed to explain the nature of the DE and the accelerated expansion including for example, quintessence (Fujii 1982; Ratra \& Peebles 1988; Chiba et al. 1997; Ferreira \& Joyce 1997,1998; Copeland 1998; Caldwell et al. 1998; Zlatev 1999), Chaplygin gas (Kamenshchik 2001), phantom energy (Caldwell 2002), k-essence (Chiba 2000; Armendariz-Picon et al. 2000,2001), tachyon (Sen 2002), ghost condensate (Arkani-Hamed et al. 2004; Piazza \& Tsujikawa 2004; Ahmed \& Moss 2008). \par

Detecting the value and evolution of the equation of state parameter $\omega = \frac{p}{\rho}$ is essential to understand the nature of DE. $\omega = 0$ for dust, $1/3$ for radiation and $-1$ for vacuum energy (cosmological constant). The varying $\omega$ which can go less than $-1$ appears in scalar field models such as phantom $\omega \leq -1$, quintessence $-1 \leq \omega \leq 1$ and quintom which is able to evolve across the cosmological constant boundary $\omega = -1$. The largest value of $\omega$ consistent with causality is $\omega=1$ for some exotic type of matter called stiff matter (Stiff fluid or Zel’dovich fluid), where the speed of sound is equal to the speed of light (Zel’dovich 1972). A matter fluid with $\omega=\frac{p}{\rho} \gg 1$ is assumed to be exist in Ekpyrotic cosmology (Khoury et al 2001; Yi-Fu Cai et al. 2012).\par
Although recent observations show the consistency of the cosmological constant DE scenario (a perfect fluid with $\omega_{\Lambda}=-1$), dynamical DE models are still allowed especially those with EoS across $-1$ (Cai, et al. 2010). The phenomenological dynamics associated with the $\omega=-1$ crossing behaviour, dubbed as quintom model, was first proposed in (Feng et al. 2005). It gives rise to the equation of state $>-1$ in the past and $<-1$ today, satisfying recent observations. The bouncing solution in a universe dominated by the Quintom matter has been studied in (Cai, et al. 2007) where it has been found that the Big Bang singularity can be avoided in the presence of the quintom matter (see Cai et al. 2010 for a comprehensive review on quintom cosmology). The simplest quintom model
can be constructed by introducing two scalar fields with one being quintessence and the other phantom. Quintom-like behavior has also been found in the context of holographic dark energy (Zhang 2006).\par 

Modified gravity theories represent another explanation approach based on modifying the geometrical part of the Einstein-Hilbert action. It can successfully explain the galactic rotation curves  with no need to DE assumption or cosmological constant (Nojiri \& Odintsov 2006,2008; Capozziello et al. 2009; De Felice and Tsujikawa 2010). Several modified gravity theories have been proposed such as $f(R)$ gravity (Carroll 2004; Capozziello 2006; Nojiri \& Odintsov 2006)where the lagrangian is a function, $f$, of the Ricci scalar
$R$, Gauss-Bonnet gravity (Nojiri \& Odintsov 2005, 2011; Cognola et al. 2008, Nojiri et al. 2008) and the torsional-based modified gravity $f(T)$ gravity (Ferraro and Fiorini 2007; Bengochea and Ferraro 2009; Linder 2010) where $f(T)$ is an arbitrary function of the torsion scalar $T$. In the framework of $f(T)$ gravity and driven by the torsion effects, the accelerated cosmic expansion can be explained without the need to introduce a new form of energy (Bengochea and Ferraro 2009; Linder 2010). It can also easily accommodate with the regular thermal expanding history (cai et al. 2016) and solve the problem of inflation with no inflaton (Ferraro and Fiorini 2007). While $f(R)$ gravity has fourth-order equations, an important advantage of $f(T)$ gravity is the second-order field equations which has attracted increasing interest in the literature.\par

Harko et al. (Harko et al. 2011), generalized $f(R)$ gravity by introducing an arbitrary function $f(R,T)$ where $R$ is the curvature scalar and $T$ is the trace of the energy momentum tensor. $f(R,T)$ gravity is an interesting version of modified gravity theories as it can reproduce the unification
of $F(R)$ and $F(T)$ gravity theories. Some cosmological aspects of $f(R,T)$ gravity have been already explored such as the thermodynamics of  FRW spacetimes (Jamil et al. 2012), the reconstruction of cosmological solutions where the late-time acceleration was obtained ( Harko et al. 2011), future singularities (Houndjo et al. 2012)و , anisotropic cosmology (Ahmed \& Pradhan 2014, Pradhan, Ahmed \& Saha 2015), scalar
perturbations (alvarenga et al. 2013). For the case of ultra–relativistic fluids, the trace of the energy momentum tensor vanishes and so they are not included in the the function $f(R,T)$. To correct this lack, a generalization has been suggested by including a new invariant, $R_{\mu\nu}T^{\mu\nu}$, in the function $f(R,T)$. This generalized model is called $f(R,T, R_{\mu\nu}T^{\mu\nu})$ gravity (Odintsov \& Saez–Gomez 2013, Hagani et al. 2013).\par

In (Ahmed \& Pradhan 2014), a specific reconstruction of $f(R,T)$ gravity has been obtained with a time varying cosmological constant given by $\Lambda(t)=-\frac{1}{2}(\rho(t)-p(t))$. This variable cosmological constant has the same expression for the negative of the thermodynamical
work density $W=\frac{1}{2}(\rho(t)-p(t))$, where $\rho$ and $p$ are the energy density and pressure of cosmic
matter, (Cai and Kim 2005;  Akbar and Cai 2007) which gives a thermodynamical meaning to the cosmological constant in this $f(R,T)$ gravity reconstruction. This specific $f(R,T)$ gravity model with varying cosmological constant has been used by many authors to study Bianchi cosmological models where a good agreement with cosmological observations has been obtained (Sahoo \& Sivakumar 2015; Bishi et al. 2017; Mahanta 2014; Chaubey and Shukla 2017; Sharma and Pradhan 2018; Shaikh and Wankhade 2017). Using this $f(R,T)$ gravity model, Sahoo et al has investigated a class of Kaluza-Klein cosmological models (sahoo et al 2016). Ram and Chandel 2014 used this gravity model to investigate Bianchi type $V$ cosmological solutions of a magnetized massive string. In this paper, we investigate the FRW cosmology of this $f(R,T)$ gravity model. \par

There is a high motivation in the literature to use variable cosmological constant which has been proposed as an attempt to solve some $\Lambda CDM$ model problems. In spite of its good agreement with observations, the $\Lambda CDM$ model suffers from several difficulties such as the fine-tuning and cosmic coincidence. Fine-tuning problem referes to the huge discrepancy between the predicted vacuum energy and the observed one (old cosmological constant problem), while cosmic coincidence means that the observed cosmological constant in the present day universe is so close to the matter density (Perlmutter et al 1999; Riess et al 1998)). Such problems have opened the door to the possible variation of cosmological constant and several cosmological scenarios with time varying cosmological constant have been proposed (Shapiro \& Sola 2009; Bonanno \& Carloni 2012; Mavromatos 2016; Pereira et al. 2017; Socorro et al. 2015; Stachowski \& Szydowski 2016). The cosmological constant represents the simplest explanation for dark energy where a very small positive energy with enough negative pressure acts as a repulsive force (repulsive gravity) that accelerates the expansion. Alternative scenarios for accelerated expansion with negative cosmological constant have been investigated in the literature where it has been shown that a negative $\Lambda$ can halt eternal acceleration, this will be discussed in details in subsection (\ref{nnn}) \par

The paper is organized as follows: section 1 is an introduction. In section 2, we review the derivation of the modified field equations with variable cosmological constant from $f(R,T)$ gravity action. In section 3, we derive the cosmological equations and utilize a form for the Hubble parameter that leads to Berman's law of constant deceleration parameter $q=m-1$ to generate a number of solutions. In section 4, we perform a detailed analysis for the solutions and investigate all possible cosmological scienarios corresponding to different valuse of $q$. The final conclusion is included in section 5. 

\section{Derivation of Field Equations} 

The $f(R,T)$ gravity action is given by (Harko et al. 2011)

\begin{equation}
S=\frac{1}{16\pi}\int{f(R,T)\sqrt{-g}d^{4}x}+\int{L_{m} \sqrt{-g}d^{4}x} , 
\end{equation}
where $L_{m}$ is the matter Lagrangian density. By varying the action $S$ with respect to $g^{\mu \nu}$, we obtain the field equations of $f(R,T)$ gravity as 

\begin{equation} \label{FieldEquations}
f_{R}(R,T)R_{\mu \nu}-\frac{1}{2} f(R,T)g_{\mu \nu}+(g_{\mu \nu} \Box  -\nabla_{\mu} \nabla_{\nu})f_{R}(R,T) 
=8\pi T_{\mu \nu}-f_{T}(R,T)T_{\mu \nu}-f_{T}(R,T)\Theta_{\mu\nu}.
\end{equation}
where $\Box = \nabla^{i}\nabla_{i}$, $f_{R}(R,T)=\frac{\partial f(R,T)}{\partial R}$, $f_{T}(R,T)=\frac{\partial f(R,T)}{\partial T}$ and $\nabla_i$ denotes the covariant derivative. $\Theta_{\mu \nu}$ and the stress-energy tensor are given by
\begin{equation}
\Theta_{\mu \nu}=-2T_{\mu\nu}+-pg_{\mu\nu},~~~~~T_{\mu\nu}=(\rho+p)u_{\mu}u_{\nu}-pg_{\mu\nu} \label{EMtensor}.
\end{equation}
where $u^{\mu}=(0,0,0,1)$ is the four velocity which satisfies $u^{\mu}u_{\mu}=1$ and $u^{\mu}\nabla_{\nu}u_{\mu}=0$. $\rho$ and $p$ are the energy density and pressure of the fluid respectively.
Different theoretical models can be obtained for each choice of $f$. Taking $f(R,T)=f_{1}(R)+f_{2}(T)$, the gravitational field equations (\ref{FieldEquations}) becomes
\begin{equation}\label{Field Equations2}
f^{'}_{1}(R)R_{\mu \nu}-\frac{1}{2} f_{1}(R)g_{\mu \nu}+(g_{\mu \nu} \Box  -\nabla_{\mu} \nabla_{\nu})f^{'}_{1}(R)= 
8\pi T_{\mu \nu}+f^{'}_{2}(T)T_{\mu \nu} +\left(f^{'}_{2}(T)p+\frac{1}{2}f_{2}(T)\right) g_{\mu\nu}. 
\end{equation}
We choose a specific form of the functions as $f_{1}(R)=\lambda_{1}R$ and $f_{2}(T)=\lambda_{1}T$ where $\lambda_{1}$ and $\lambda_{2}$ are arbitrary parameters. Taking $\lambda_{1}=\lambda_{2}=\lambda$ leads to $f(R,T)=\lambda (R+T)$ and then equation (\ref{Field Equations2}) can be written as 
\begin{equation}
\lambda R_{\mu \nu}-\frac{1}{2} \lambda (R+T)g_{\mu \nu}+(g_{\mu \nu} \Box  -\nabla_{\mu} \nabla_{\nu})\lambda 
=8\pi T_{\mu \nu}-\lambda T_{\mu \nu}+\lambda (2T_{\mu \nu}+pg_{\mu \nu}). \label{Fsubst}
\end{equation}
setting $(g_{\mu \nu} \Box  -\nabla_{\mu} \nabla_{\nu})\lambda=0$ and after simple rearrangement we get
\begin{equation}
G_{\mu\nu}-\left(p +\frac{1}{2} T\right) g_{\mu\nu}=\frac{8\pi+\lambda}{\lambda} T_{\mu \nu}.    \label{ours}
\end{equation}
where $G_{\mu\nu}=R_{\mu\nu}-\frac{1}{2}g_{\mu\nu}R$ is the Einstein tensor. Recalling Einstein equations with cosmological constant
\begin{equation}
G_{\mu\nu}+\Lambda g_{\mu\nu}=8\pi T_{\mu \nu}. \label{Einst}
\end{equation}
To ensure the same sign of the RHS of (\ref{Einst}) and (\ref{ours}), the inequality $\frac{8\pi+\lambda}{\lambda}>0$ must be satisfied which implies that $\lambda>0$ or $\lambda<-8\pi$. The term $\left(p +\frac{1}{2} T\right)$ can now be considered as the negative of the cosmological constant. i.e, $p +\frac{1}{2} T \equiv -\Lambda(T)$. So, for the energy momentum tensor (\ref{EMtensor}) we get
\begin{equation}
\Lambda(t)=-\frac{1}{2}(\rho(t)-p(t)) \label{cosm}.
\end{equation} 
Taking into account the general equation of state $p=\omega \rho$, then $\Lambda(t)$ could also be expressed as $\Lambda(t)=\frac{1}{2}\rho(t)(\omega-1) \label{cosm2}$.

\section{Cosmological equations}
 
The metric that describes a homogeneous and isotropic universe is the FRW metric given by

\begin{equation}
ds^{2}=-dt^{2}+a^{2}(t)\left[ \frac{dr^{2}}{1-Kr^2}+r^2d\theta^2+r^2\sin^2\theta d\phi^2 \right] \label{RW}
\end{equation} 
where $r$, $\theta$, $\phi$ are comoving spatial coordinates, $a(t)$ is the cosmic scale factor, $t$ is time, $K$ is either $0$, $-1$ or $+1$ for flat, open and closed universe respectively. Applying equation (\ref{ours}) to the metric (\ref{RW}) we get the following cosmological equations

\begin{eqnarray}
H^2 &=& \frac{8\pi+\lambda}{3\lambda} \rho-\frac{1}{3}\Lambda(t)-\frac{
K}{a^2} \label{RW2}.\\
 \frac{\ddot{a}}{a} &=& -\frac{8\pi+\lambda}{2\lambda}(\rho+p)   \label{RW1}.
\end{eqnarray}

The field equations (\ref{RW1}) and (\ref{RW2}) contain three unknown parameters $a$, $p$ and $\rho$.
In ( Maharaj \& Naidoo 1993), the generalized Einstein equations with variable cosmological constant $\Lambda(t)$ has been considered for FRW metric where the following variation of the Hubble parameter has been assumed:
\begin{equation} \label{hub}
H=Da^{-m}.
\end{equation}
Where $D$ and $m$ are constants. This Hubble parameter form (\ref{hub}) was first utilised in (Berman 1983; Berman \& Gomide 1988) and it leads to Berman's law of constant deceleration parameter $q = m - 1,~m\geq 0$. In ( Maharaj \& Naidoo 1993), a number of classes of new solutions for variable cosmological constant $\Lambda(t)$ has been introduced:
\begin{align} \label{scale1}
   a = \left\{
     \begin{array}{lr}
       \left(C+mDt\right)^{\frac{1}{m}}~~~~~~\mbox{for}~ m\neq 0,\\
       Ee^{Dt} ~~~~~~~~~~~~~~~~~\mbox{for}~ m= 0.
     \end{array}
   \right.
\end{align} 
In order to solve the modified field equations (\ref{ours}) with its time varying cosmological constant $\Lambda(t)=\frac{1}{2}\left(\rho(t)-p(t)\right)$, we are going to make use of (\ref{scale1}) and investigate different values of $m$ corresponding to a decelerating expansion where $q = m - 1 > 0$ (i.e. $m>1$) and an accelerating expansion where $q = m - 1 < 0$ (i.e. $m<1$).

\section{Solutions}
\subsection{The case for $q>0$ ($m>1$): A decelerating radiation-dominated flat universe }
For $m\neq 0$, the metric (\ref{RW}) has the form
\begin{equation}
ds^{2}=-dt^{2}+\left(C+mDt\right)^{\frac{1}{m}}\left[ \frac{dr^{2}}{1-Kr^2}+r^2d\theta^2+r^2\sin^2\theta d\phi^2 \right] 
\end{equation} 
Substituting (\ref{scale1}) in (\ref{RW1}) and (\ref{RW2}) we get the pressure $p$ and energy density $\rho$ as
\begin{equation}\label{p}
p=72\lambda \:\frac{K\left(\pi+\frac{\lambda}{6}\right)\left(Amt+C\right)^{\frac{2m-2}{m}}-\frac{2}{3}mA^2\left(\pi+\frac{\lambda}{16}\right)+  \frac{5}{3}A^2\left(\pi+\frac{\lambda}{8}\right)}{(192\pi^2+40\pi\lambda+3\lambda^2)\left(Amt+C\right)^{2}}.
\end{equation} 
\begin{equation}\label{rho}
\rho=3\lambda \frac{24K\left(\pi+\frac{\lambda}{12}\right)\left(Amt+C\right)^{\frac{2m-2}{m}}+\left((m+1)\lambda+24\pi\right)A^2}{(192\pi^2+40\pi\lambda+3\lambda^2)\left(Amt+C\right)^{2}}.
\end{equation} 
Using (\ref{p}) and (\ref{rho}) we obtain the equation of state parameter $\omega$ as:
\begin{equation}
\omega=\frac{-16A^2\left(\left(\pi+\frac{\lambda}{16}\right)m-\frac{5}{2}\pi-\frac{5}{16}\lambda\right)\left(Amt+C\right)^{\frac{2}{m}}
+24K\left(\pi+\frac{\lambda}{6}\right)\left(Amt+C\right)^{2}}{24K\left(\pi+\frac{\lambda}{12}\right)\left(Amt+C\right)^{2}+\left((m+1)\lambda+24\pi\right)A^2}.
\end{equation} 
The cosmological constant (\ref{cosm}) is given by
\begin{equation}
\Lambda=-3\lambda\frac{K\lambda \left(Amt+C\right)^{\frac{2m-2}{m}}+\left(8\pi+2\lambda-m(8\pi+\lambda)\right)A^2}{(192\pi^2+40\pi\lambda+3\lambda^2)\left(Amt+C\right)^{2}}.
\end{equation} 

Figure 1 (a), (b) and (c) shows variation of pressure $p$, density $\rho$ and cosmological constant $\Lambda$ versus time. The pressure is a positive decreasing function for flat and closed universes, it starts from a large positive value and then approaches a small positive value near zero. The energy density is negative for $K=-1$ and hence the open universe is not possible. For flat and closed universes, the energy density is positive and tends to zero as the cosmic time $t \rightarrow \infty$. The cosmological constant $\Lambda$ (figure 1 (c)) is positive, it reaches a very small positive value at the current epoch which agrees with observations (Perlmutter et al. 1999; Riess et al. 1998; Tonry et al. 2003; Clocchiatti et al 2006). The equation of state parameter (Figure 1 (i)) $\omega=0.3335984859 \approx \frac{1}{3}$ for $K=0$ which means a universe very close to the radiation-dominated era where the cosmic expansion was decelerating (Liddle 2003). It is generally believed that the universe is decelerating during the radiation and dark matter dominated eras, and is accelerating during the early and late-time inflations eras (Perico et al. 2013). The decelerated radiation-dominated era in this model arises with $q=1$ in a good agreement with the complete cosmic history investigated in (Perico et al. 2013). For $K=1$, $\omega \rightarrow 1$ which means a decelerating universe approaching a stiff matter-dominated era where $\omega=1$. Table (\ref{tap}) shows the behavior of $p(t)$, $\rho(t)$, $\Lambda(t)$, $c_s^2$ and energy conditions verses cosmic time for different values of $m$.

\begin{table}[H]\label{tap}
\centering
\tiny
    \begin{tabular}{ | p{1.1cm} | p{2cm} | p{2cm} | p{2cm} | p{2cm} | p{2cm} | p{2cm} |}
    \hline
           & $m=2$ & $m=1$ & $m=\frac{1}{2}$ & $m=\frac{2}{3}$ & $m=-2$ & $m=0$\\ \hline
    p & +ve for $K=0,1$ & +ve for all $K$  &  +ve for all $K$  & +ve for all $K$  & $\infty$ for all $K$ & +ve for $K= 0$  \\ \hline
    $\rho$ & +ve for $K=0,1$ & +ve for all $K$  &  +ve for all $K$  & +ve for all $K$  & $\infty$ for all $K$ & +ve for $K= 0$ \\ \hline
    $\Lambda$ & +ve for all $K$ & -ve for all $K$  &  -ve for all $K$  & -ve for all $K$  & -$\infty$ for all $K$ & -ve for all $K$  \\ \hline
		$c_s^2$ & valid for $K=0$ & invalid for all $K$  & invalid for all $K$ & invalid for all $K$  &  invalid for all $K$& valid for all $K$ \\ \hline
		$p+\rho$ & valid for $K=0,1$ & +ve for all $K$  &  +ve for all $K$  & +ve for all $K$  & invalid for all $K$ & +ve for $K= 0$  \\ \hline
		$\rho+3p$ & valid for $K=0,1$ & +ve for all $K$  &  +ve for all $K$  & +ve for all $K$  & invalid for all $K$  & +ve for $K= 0$  \\ \hline
		$\rho-p$ & valid for all $K$ & -ve for all $K$  &  +ve for all $K$  & +ve for $K=0$ & invalid for all $K$  & +ve for $K= -1$  \\ \hline
    \end{tabular}
		\caption {The behavior of $p$, $\rho$, $\Lambda$, $c_s^2$ and energy conditions verses cosmic time for different $m$.}
		\end{table}
		
\subsection{Stability of solutions} 

The physical acceptability of the model can be checked through testing the energy conditions and the sound speed. The null, weak, strong and dominant energy conditions are respectively given by (Hawking \& Ellis 1973; Wald 1984): $\rho + p \geq 0$; $\rho \geq 0$, $\rho + p \geq 0$; $\rho + 3p \geq 0$ and $\rho \geq \left|p\right|$. The energy conditions have been plotted in figure 1 (d), (e) and (f). They are all satisfied for flat and closed universes. Now, the adiabatic square sound speed for any fluid is defined as: $c_s^2=\frac{dp}{d\rho}
$. In addition to the positivity of $c_s^2$, the causality condition must be satisfied too. Causality implies that the sound speed must not exceed the speed of light. Since we are using relativistic units in which $c=G=1$, the sound speed should exist within the range $0 \leq \frac{dp}{d\rho} \leq 1$. For the current model, we get the sound speed as
\begin{equation}
c_s^2=\frac{-16mA^2\left(\left(\pi+\frac{\lambda}{16}\right)m-\frac{5}{2}\pi-\frac{5}{16}\lambda\right)\left(Amt+C\right)^{\frac{2}{m}}
+24K\left(\pi+\frac{\lambda}{6}\right)\left(Amt+C\right)^{2}}{mA^2(m\lambda+24\pi+\lambda)\left(Amt+C\right)^{\frac{2}{m}}+24K\left(\pi+\frac{\lambda}{12}\right)\left(Amt+C\right)^{2}}
\end{equation}
Figure 1 (g) shows that this quantity is always less than one for $K=0$ and tends to 1 for $K=1$. Again, for the stiff matter dominated era the sound speed is equal to the speed of light. So, in summary, $K=0$ describes a stable flat decelerating universe very close to the radiation dominated era ($\omega \approx \frac{1}{3}$), While $K=1$ describes a stable closed decelerating universe approaching the stiff matter dominated era ($\omega \approx 1$ and $c_s^2 \approx 1$). While all figures have been plotted with $\lambda=0.01$, we have tried other values for $\lambda$ in figure 1 (h) for the flat $K=0$ case supported by observations. We have found that $c_s^2$ is always $\leq 1$ for any positive value of $\lambda$. As $\lambda$ increases, $c_s^2$ increases and tends to $1$ for very large values of $\lambda$. For the possible negative values of $\lambda<-8\pi$, we found that $c_s^2$ is always $\geq 1$ which means no stability for negative $\lambda$.

\subsection{The case for $q<0$ ($m<1$): A universe with negative cosmological constant.}\label{nnn}
\subsubsection{$m=0$ ($q=-1$).}
For $m=0$, the metric (\ref{RW}) is
\begin{equation}
ds^{2}=-dt^{2}+Ee^{Dt}\left[ \frac{dr^{2}}{1-Kr^2}+r^2d\theta^2+r^2\sin^2\theta d\phi^2 \right] 
\end{equation} 
The expressions for $p$, $\rho$, $\Lambda$ and $\omega$ are

\begin{equation}
p= \frac{120\lambda\left(\frac{3}{5}\left(\lambda+\frac{\pi}{6}\right)Ke^{-2Mt}+E^2M^2\left(\pi+\frac{\lambda}{8}\right)\right)}{E^2\left(192\pi^2+40\pi \lambda+3\lambda^2\right)}
\end{equation} 

\begin{equation}
\rho= \frac{72\lambda\left(K\left(\pi+\frac{\lambda}{12}\right)Ke^{-2Mt}+E^2M^2\left(\pi+\frac{\lambda}{24}\right)\right)}{E^2\left(192\pi^2+40\pi \lambda+3\lambda^2\right)}
\end{equation} 

\begin{equation}
\Lambda=-\frac{24\lambda\left(\frac{1}{8}\lambda Ke^{-2Mt}+E^2M^2\left(\pi+\frac{\lambda}{4}\right)\right)}{E^2\left(192\pi^2+40\pi \lambda+3\lambda^2\right)}
\end{equation} 

\begin{equation}
\omega=\frac{24K\left(\pi+\frac{\lambda}{6}\right)e^{-2Mt}+40E^2M^2\left(\pi+\frac{\lambda}{8}\right)}{24K\left(\pi+\frac{\lambda}{12}\right)e^{-2Mt}+24E^2M^2\left(\pi+\frac{\lambda}{24}\right)}
\end{equation} 

In this case, we have a stable flat universe with negative cosmological constant, and positive energy density (table \ref{tap}). Only the spatially flat universe is allowed (as observations tell us) where the closed and open universes have negative energy density. Although recent observations suggest a positive cosmological constant $\Lambda>0$, a negative cosmological constant which can fit quite well a large data set is also possible and can solve the eternal acceleration problem (Vincenzo et al. 2008). The flat and accelerating universe, confirmed by observations, doesn't have to keep accelerating forever and this eternal acceleration is a consequence of assuming a positive cosmological constant to explain the accelerated expansion (Vincenzo et al. 2008). This radical different approach of negative cosmological constant has been investigated by many authors (Gardenas et al. 2003, Grande et al. 2006, Gu and Hwang 2006, Vincenzo et al. 2008, Prokopec 2011, Landry et al. 2012, Kayll Lake , Maeda and Ohta 2014, Baier et al. 2015 ,Chruściel et al. 2018). The AdS/CFT correspondence presents a strong argument that the negative $\Lambda$ case should also be examined (Aharony et al. 2000). Prokopec 2011, introduced modifications to Friedmann cosmology that can lead to observationally viable cosmologies with $\Lambda<0$. Maeda and Ohta in 2014 studied gravitational theories with a cosmological constant and the Gauss-Bonnet curvature squared term and found stable de Sitter solution with negative $\Lambda$. Landry et al. 2012, studied the McVittie solution with $\Lambda<0$ and found that the negative $\Lambda$ case ensures collapse to a Big Crunch as in pure FRW case. Chruściel et al. 2018, proved existence of large families of solutions of Einstein-complex scalar field equations with $\Lambda<0$. So, the accelerating solution we have got for $m=0$ ($q=-1$) with negative $\Lambda$ is interesting and has been an active research point.

\subsubsection{Other values for $m<1$.}

We have investigated other values for $m<1$ as shown in table (\ref{tapp}). Wile we have $\Lambda<0$ and $\rho>0$ for all $K$, the stability of the solutions is weaker than the $m=0$ case in which the only flat universe is possible.

\begin{table}[H]\label{tapp}
\centering
\tiny
    \begin{tabular}{ | p{1.1cm} | p{2cm} | p{2cm} | p{2cm} | p{2cm} | p{2cm} | p{2cm} |}
    \hline
           & $m=0.1$ & $m=0.2$ & $m=0.4$ & $m=0.6$ & $m=0.7$ & $m=0.8$\\ \hline
    p & +ve for all $K$ & +ve for all $K$  &  +ve for all $K$  & +ve for all $K$  &  +ve for all $K$  & +ve for all $K$  \\ \hline
    $\rho$ & +ve for all $K$ & +ve for all $K$  &  +ve for all $K$  & +ve for all $K$  &  +ve for all $K$  & +ve for all $K$ \\ \hline
    $\Lambda$ & -ve for all $K$ & -ve for all $K$  &  -ve for all $K$  & -ve for all $K$  &  -ve for all $K$  & -ve for all $K$  \\ \hline
		$c_s^2$ & invalid for all $K$ & invalid for all $K$  & invalid for all $K$ & invalid for all $K$  &  invalid for all $K$& invalid for all $K$ \\ \hline
		$p+\rho$ & +ve for all $K$ & +ve for all $K$  &  +ve for all $K$  & +ve for all $K$ & +ve for all $K$ &  +ve for all $K$  \\ \hline
		$\rho+3p$ & +ve for all $K$ & +ve for all $K$  &  +ve for all $K$  & +ve for all $K$ & +ve for all $K$ &  +ve for all $K$  \\ \hline
		$\rho-p$ & -ve for all $K$ & -ve for all $K$  &  -ve for all $K$  & -ve for all $K$ & -ve for all $K$ &  -ve for all $K$  \\ \hline
    \end{tabular}
		\caption {The behavior of $p$, $\rho$, $\Lambda$, $c_s^2$ and energy conditions verses cosmic time for different values of $0<m<1$.}
		\end{table}
		
\section{Conclusion}

We have constructed a general FRW cosmological model in $f(R,T)$ gravity reconstruction with variable cosmological constant. A specific form for the Hubble parameter that leads to Berman's law $q=m-1$ has been utilized which generated a number of solutions to the modified Friedmann equations. The possibility and stability of decelerating ($q>0$) and accelerating ($q<0$) solutions have been investigated. For the decelerating case, we found the most stable solution is a decelerating radiation-dominated flat universe at $q=1$ in a good agreement with cosmic history in the literature. For the accelerating case, the most stable solution is a flat universe with positive energy density and negative cosmological constant at $q=-1$. We have discussed the possibility of several stable accelerating solutions with $\Lambda<0$ in the literature in which the eternal acceleration problem disappears. Nonconventional mechanisms that are expected to address the late-time acceleration has been an active research area and it is commonly acknowledged that the early universe application would be nontrivial as well. For example, a nonsingular bouncing solution might arise in these models (Cai 2014). In the context of inflationary cosmology, there are several conceptual challenges without solutions which provide a good motivation to search for alternative proposals describing the very early universe beyond the standard inflationary $\Lambda CDM$ model. In particular, we mention two interesting phenomenological scenarios alternative to the inflationary $\Lambda CDM$ model which are the healthy nonsingular bouncing cosmology (Cai et al. 2012) and the matter-bounce in inflationary scenario (Cai et al. 2009). In the matter-bounce inflationary scenario, the inflationary cosmology is generalized by introducing a matter-like contracting phase before the inflationary phase. This scenario was found to be very powerful in reconciling the tension between the Planck and BICEP2 observations when compared with the $\Lambda CDM$ (Xia et al. 2014). In bouncing cosmology, the expansion of the universe is preceded by an initial phase of contraction with a bouncing point connecting the contraction and the expansion. A comprehensive introduction has been given in (Cai 2014).

\section*{Acknowledgment}
We are so grateful to the reviewer for his many valuable suggestions and comments that significantly improved the paper.

\begin{figure}[H]
  \centering
  \subfigure[$p$]{\label{fig:rrf}\includegraphics[width=0.27\textwidth]{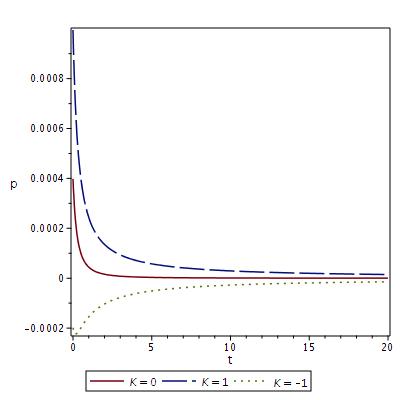}}                
  \subfigure[$\rho$]{\label{f222}\includegraphics[width=0.27\textwidth]{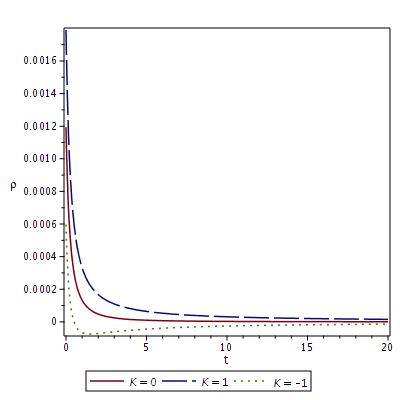}}
  \subfigure[$\Lambda$]{\label{F333}\includegraphics[width=0.27\textwidth]{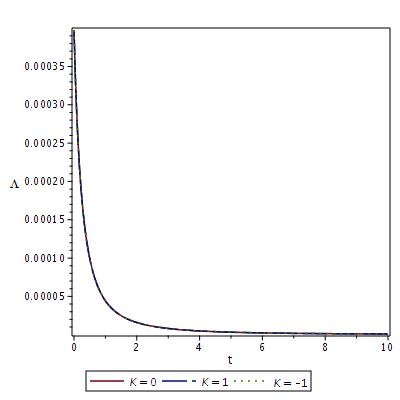}}\\             
  \subfigure[$\rho+p$]{\label{F555}\includegraphics[width=0.27\textwidth]{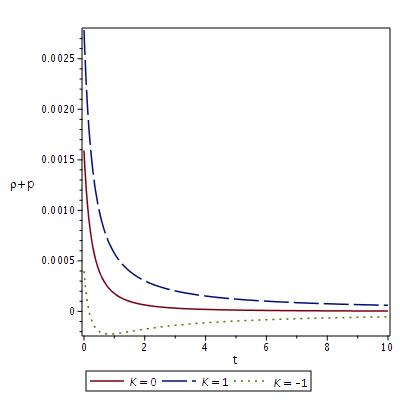}}
  \subfigure[$\rho+3p$]{\label{F63}\includegraphics[width=0.27\textwidth]{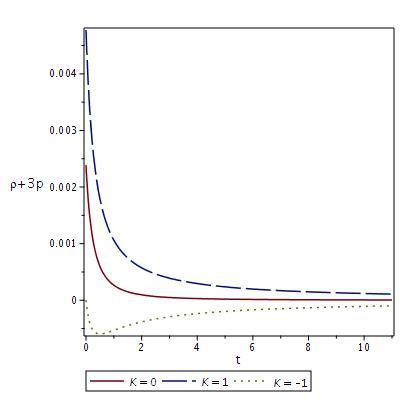}} 
	\subfigure[$\rho-p$]{\label{F423}\includegraphics[width=0.27\textwidth]{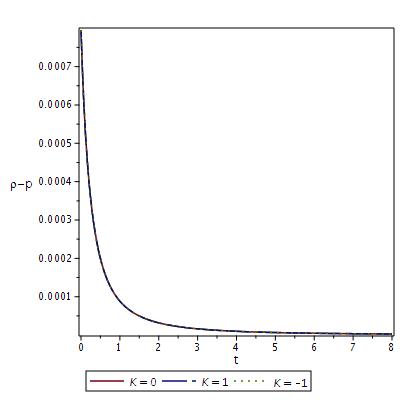}} \\
	\subfigure[$c_s^2$]{\label{F67}\includegraphics[width=0.27\textwidth]{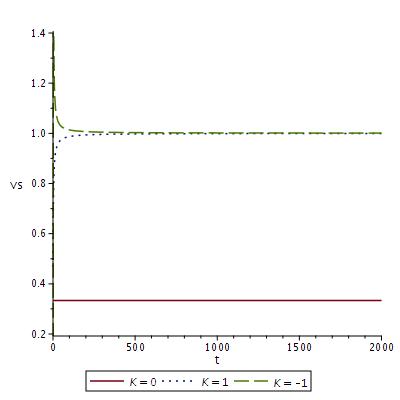}} 
	\subfigure[$c_s^2$ for $K=0$.]{\label{F678}\includegraphics[width=0.27\textwidth]{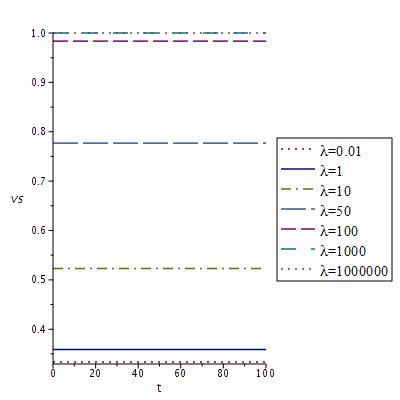}}
	\subfigure[$\omega$]{\label{F68}\includegraphics[width=0.27\textwidth]{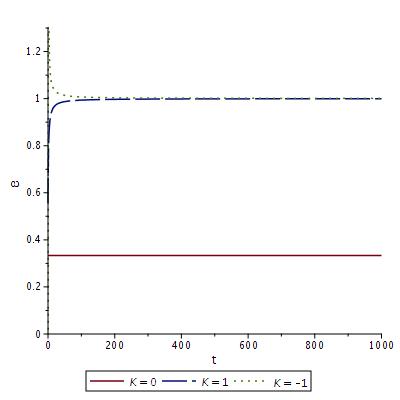}} 
  \caption{The behavior of $p$, $\rho$, $\Lambda$,$\omega$, $c_s^2$ and energy conditions verses cosmic time for $m=2$. Here $A=C=2$ and $\lambda=0.01$. $\omega \approx \frac{1}{3}$ for flat universe.}
  \label{fig:cassimir55}
\end{figure}

\end{document}